\documentclass[onecolumn]{aastex701}

\usepackage{amsmath}
\usepackage{xcolor}
\usepackage{ulem}
\makeatletter
\renewenvironment{acknowledgments}{
  \section*{Acknowledgments}
}

\usepackage{xspace}
\usepackage{multirow}
\submitjournal{ApJ}

\shorttitle{Multi-Path QFP Magnetoacoustic Waves}
\shortauthors{Miao et al.}

\newcommand{\secref}[1]{Section \ref{#1}}
\newcommand{\figref}[1]{Figure~\ref{#1}}
\newcommand{\kms}{\ensuremath{\,\mathrm{km}\cdot \mathrm{s}^{-1}}}
\newcommand{\unit}[1]{\ensuremath{\,\mathrm{#1}}}
\newcommand{\degree}{\ensuremath{^\circ}}

\newcommand{\speed}[1]{#1 km$\cdot$s${}^{-1}$}
\usepackage{lineno}
\begin{document}
\nolinenumbers

\title{Multi-Path Quasi-Periodic Fast-mode Propagating Magnetoacoustic Waves to Diagnose Coronal Magnetic Field and Flaring Core}
\author[orcid=0000-0003-2183-2095,sname='Miao']{Yuhu Miao}
\affiliation{School of Information and Communication, Shenzhen University of Information Technology, Shenzhen 518172, China}
\email[show]{miaoyh@suit-sz.edu.cn} 

\begin{abstract}

Quasi-periodic fast-mode magnetoacoustic waves are often detected during solar flare events, although they are not observed in every flare, due to observational signal-to-noise limits and differences in flare magnetic topology and energy release strength. These structures propagate along magnetic configurations and supply effective diagnostics for coronal magnetic environments and flaring regions. Periodic signatures seen in fast-mode QFP wave trains carry physical information about excitation processes and propagation conditions. These signatures support quantitative studies of flare cores and magnetic channel properties. This work focuses on a well-documented event involving two oppositely oriented QFP waves simultaneously excited by a GOES-class M6.0 solar flare that occurred in active region NOAA 11261 on August 3, 2011. These QFP waves can be categorized into broad and narrow wave trains, with the narrow one propagating along funnel-like loops and the broad one moving through the low corona. Observational results suggest that both broad-wave and narrow-wave QFP phenomena can be simultaneously triggered by a single flare eruption. This study also indicates that such multi-path QFP wave events can be utilized to diagnose the magnetic field and the flare's core, and demonstrates the capability of multi-path QFP waves for robust coronal magnetic field and flare core diagnostics.

\end{abstract}

\section{Introduction}

Quasi-periodic fast-mode propagating (QFP) waves constitute a key component of solar magnetohydrodynamic activity. These waves shape many energetic processes in the solar atmosphere. QFP waves, as a frequently observed phenomenon, have been investigated in a series of reports \citep{liuw11,liuw12,2012ApJ...753..112Z,shenyd12b,shenyd2013b,yuanding2013,warmuth2015,tianhui2016,kumar2017,shenyd2018,miao2019,miao2020,yuand2023,zhongsh2023,kolotkov2023,wangtj2026}.
These fast-mode waves exhibit rapid propagation and quasi-periodic modulation. The fast-mode waves can travel along or at an angle to the magnetic field \citep{dingmd2001,nakariakov2005,wangyuming2002,tianhui2012,shen2017,miao2026raa,miao2026universe}.

Observational evidence of QFP waves has been obtained from various solar imaging instruments, including the Atmospheric Imaging Assembly (AIA) on board the Solar Dynamics Observatory (SDO) \citep{liuw11,ofman2011,ofman2025,wangtj2026} and the Extreme-ultraviolet Imaging Telescope (EIT) on the Solar and Heliospheric Observatory (SOHO) \citep{delaboudiniere1995,thompson1998,warmuth2007}. High-resolution observations have detected these waves in a range of solar structures, such as active regions, prominences, coronal loops \citep{wangtj2021,2024PrA....42..429M}, and coronal holes \citep{chenpf2011}. Their association with phenomena like solar flares \citep{2013MNRAS.431.1359Z,guoyang2015,zhangqm2022,miao2025pof}, filament eruptions \citep{chenhechao2025}, and coronal mass ejections highlights their significance in understanding the Sun's eruptive behavior \citep{chenyao2010}.

\citet{takasao2016} conducted numerical magnetic reconnection experiments, demonstrating how reverse plasma flows originating in reconnection regions form a magnetic tuning fork above the flare loop apex. Acting as an Alfvénic resonator, this region serves as the source of quasi-periodic processes, from which QFP waves may potentially leak from the loop’s apex. This scenario was further validated by multi-temperature observations from SDO/AIA. Additionally, \citet{goddard2019} conducted a parametric study on the initial impulsive driving force of QFP waves. They found that the final spatial and spectral characteristics of the guided QFP wave trains are strongly influenced by the duration of the initial disturbance. This dependency highlights the potential of utilizing excited QFP waves for diagnosing the flare core region \citep{takasao2016,lileping2018,miao2021a,shenyd2022,2025PhFl...37c7177O}.

Numerical simulations and theoretical models have offered valuable insights into the generation, propagation, and dissipation mechanisms of QFP waves. Magnetohydrodynamic (MHD) simulations that incorporate realistic solar magnetic field configurations have successfully reproduced observed wave phenomena and clarified the physical processes driving their dynamics \citep{ofman2011,ofman2018,ofman2025,wangtj2026}. Additionally, theoretical studies have investigated the influence of various parameters, including magnetic field strength, plasma density, and temperature, to further improve knowledge of wave behaviors. The emergence of QFP wave trains is probably triggered by abrupt bursts of flaring energy, and it was suggested that QFP wave trains could be spontaneously stimulated and transition into a quasi-periodic state under the guiding influence of magnetized coronal structures\citep{yuanding2013,miao2021a,shenyd2022}. This phenomenon was exemplified by injecting impulsive energy into a divergent magnetic funnel. The quasi-periodic characteristics were accurately replicated in both the confined fast magnetoacoustic waves and the leaking mode\citep{pascoe2013, pascoe2014b,quzhining2017}. Numerous studies have corroborated the dispersive progression of fast magnetoacoustic waves within coronal waveguides\citep{nistico2014,shenyd2018,miao2019,miao2020}.

Despite significant progress in observational and numerical studies, several key questions regarding QFP waves remain unresolved. The exact mechanisms driving their generation and energy transport, the influence of magnetic topology on their characteristics, and their potential impact on solar and space weather phenomena are still subjects of active research and debate within the solar physics community. \citet{ofman2018} reported the first observation of counter-propagating QFP waves and provided the 3D MHD model of this event. Recently, \citet{miao2021a} reported a bidirectional quasiperiodic fast-mode propagating (BQFP) magnetoacoustic wave with full-disk imaging capability of the Atmospheric Imaging Assembly \citep[AIA;][]{lemen12} onboard the \textit{Solar Dynamics Observatory} \citep[\textit{SDO};][]{pesnell12}. The authors demonstrated that BQFP waves serve as an effective tool to probe the evolution of flaring core and the features of magnetic field of the coronal loops (waveguide). \citet{zhouxp2024bqfp} observed both a broad QFP wave and a narrow QFP wave trains within the same event in their simulations. The two BQFP events are closely associated with the energy release of the flaring core \citep{2024PrA....42..224M}.

Coronal seismology is a powerful diagnostic tool for measuring the physical parameters of coronal plasmas, which are difficult to obtain directly. This technique has been applied to a wide range of coronal structures. For instance, seismology of streamer waves has been used to constrain the magnetic field in large-scale coronal streamers, and this represents only one of many applications of the method. \citet{chenyao2011} established a coronal seismological method using streamer waves to measure the magnetic field in the outer corona. They inferred radial magnetic field profiles from wave phase speeds, solar wind velocities, and electron densities, and revealed the temporal decay of the magnetic field during the post-CME coronal recovery. Similarly, \citet{kwon2013} applied streamer-wave seismology to derive the global coronal magnetic field with STEREO observations.
Other applications of coronal seismology include measurements in coronal loops, active region funnels, and flaring core regions, providing independent constraints on the coronal magnetic field and plasma properties. This work focuses on local, high-precision magnetic field diagnostics using fast-mode QFP waves, whereas the streamer-wave approach targets large-scale global magnetic field structures. Both methods are fundamentally important and play complementary roles in coronal magnetic field diagnostics. The former is well suited for probing small-scale local magnetic field variations, while the latter performs better in deriving global coronal magnetic field distributions. These two distinct approaches thus provide comprehensive and complementary constraints for coronal magnetic field diagnostics.

The origin of periodic signals in multi-path QFP fast-mode waves from a flare is identified in the present analysis. Narrow and broad fast-mode wave trains show consistent periodic signals. These signals match the periodicity measured in flare energy release. It should be noted that \citet{liuw11} provided the first observational evidence for the physical link between flare periodicity and QFP wave periodicity, using nonthermal hard X-ray emissions. The potential of bidirectional QFP waves for constraining the origin of periodicity and investigating plasma diagnostic capabilities is explored. In \secref{sec:instr}, the flare-triggered QFP event and relevant data analysis are described. \secref{sec:obs} presents the main results; the discussion and conclusions are given in \secref{sec:sum}.

\section{Observations and Data analysis}
\label{sec:instr}

\begin{figure}[ht]
	\epsscale{1.2} \plotone{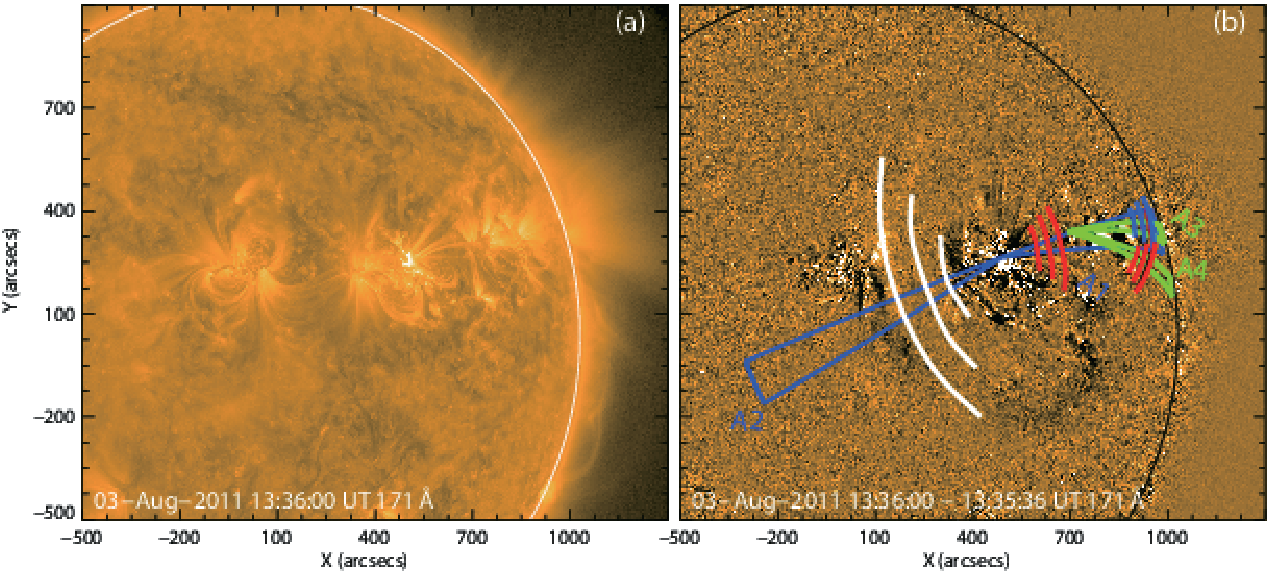}
	\caption{(a) SDO/AIA 171 \AA\ field-of-view snapshot of NOAA 11261 at 13:36 UT on 03 August 2011, showing funnel-shaped coronal magnetic waveguides. (b) AIA 171 \AA\ running-difference image showing bidirectional QFP wave trains excited by the M6.0 flare. White arcs denote the broad low-coronal wave train; red and blue arcs mark narrow funnel-guided QFP wave trains that split into two sub-trains near 13:45 UT. Sectors A1 - A4 correspond to the cuts used for time-distance analysis in \figref{fig:fig2}{}. An H.264 MP4 animation available in the online HTML version compresses observations from 12:59 - 14:14 UT (75 min real observing time) into a 7 s playback clip built from 12 s cadence AIA images. The animation tracks the full dynamic evolution of flare-driven QFP waves, including concurrent generation of oppositely directed narrow and broad wave populations, outward spread of the diffuse broad wave, and splitting of the narrow funnel wave trains at ~13:45 UT. This splitting phenomenon is only visible in the time-continuous animation and cannot be captured by the static panels printed here. This descriptive text supports offline and visually impaired readers unable to view the streaming video.\label{fig:fig1}}
\end{figure}

\begin{figure}[ht]
	\epsscale{1.2} \plotone{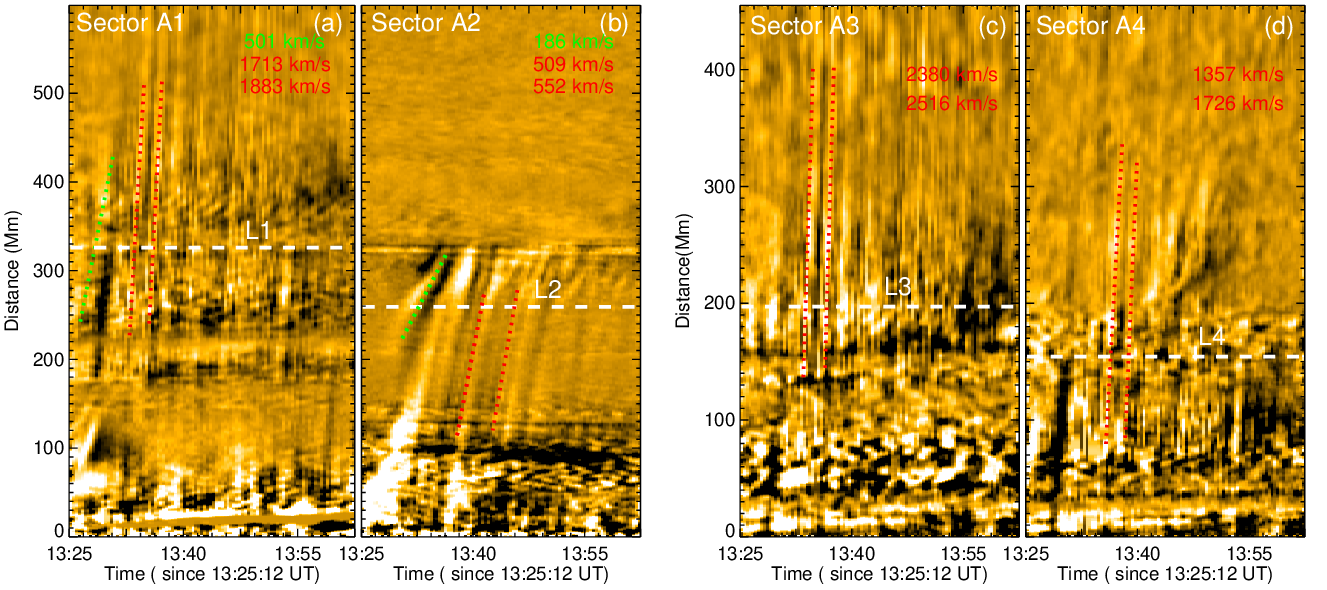}
	\caption{SDO/AIA running difference images and corresponding time–distance diagrams for sectors ``A1–A4''. White dashed lines indicate positions for wavelet analysis (L1–L4). Green and red dotted lines trace typical QFP wave fronts, whose slopes yield propagation speeds.\label{fig:fig2}}
\end{figure}

\begin{figure}[ht]
	\epsscale{1.0} \plotone{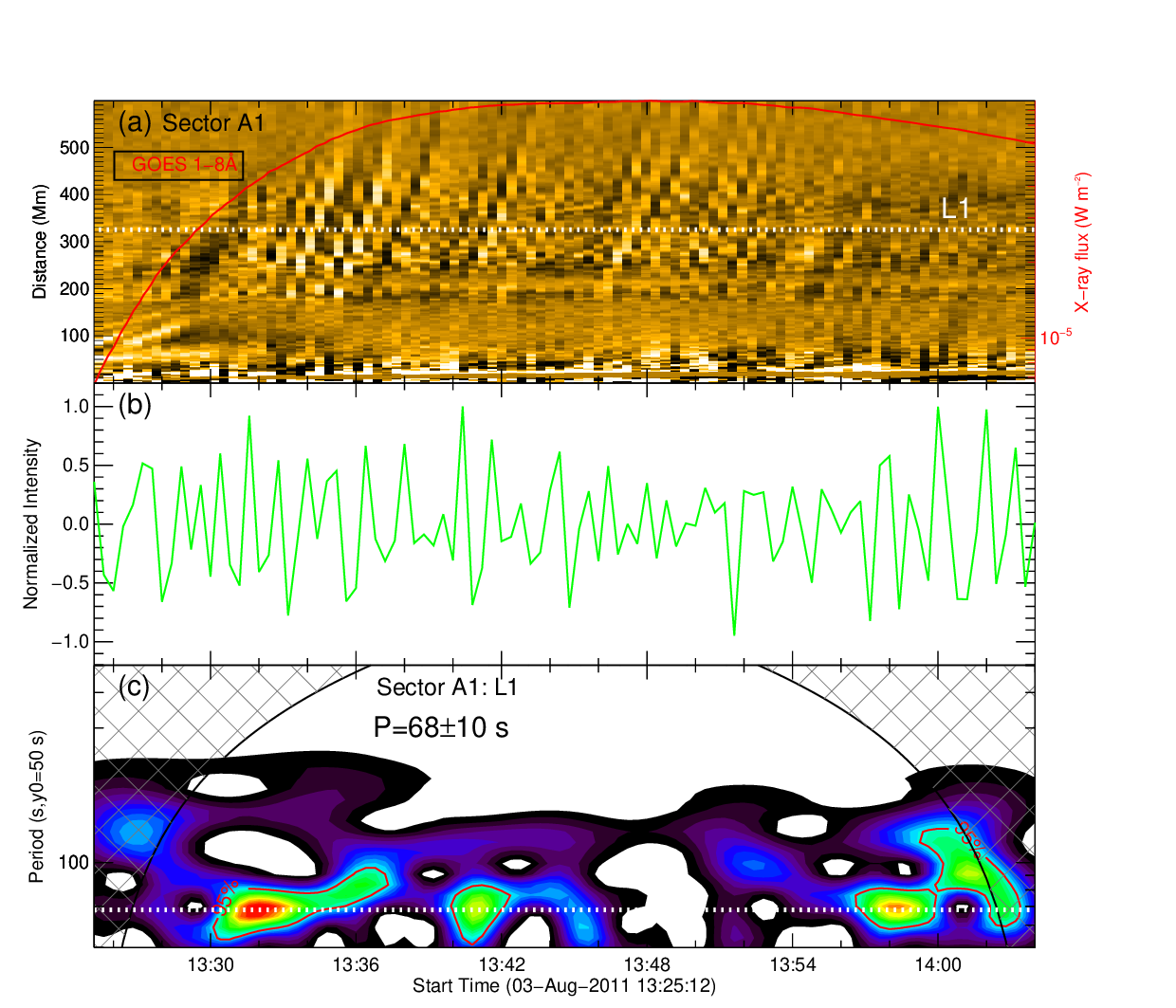}

\caption{ (a) Time-distance plot of Sector ``A1'', overlaid with the GOES X-ray flux (red curve). (b) Detrended emission intensity measured at L1 in \figref{fig:fig2}(a).
(c) Wavelet spectrum derived from the L1 light curve displayed in panel (b), where the dominant periodic signal is measured to be $68\pm10$ s. The symbol $y_0$ denotes the initial value of the y-axis in panel (c).}
\label{fig:fig3}
\end{figure}

\begin{figure}
	\epsscale{1.2} \plotone{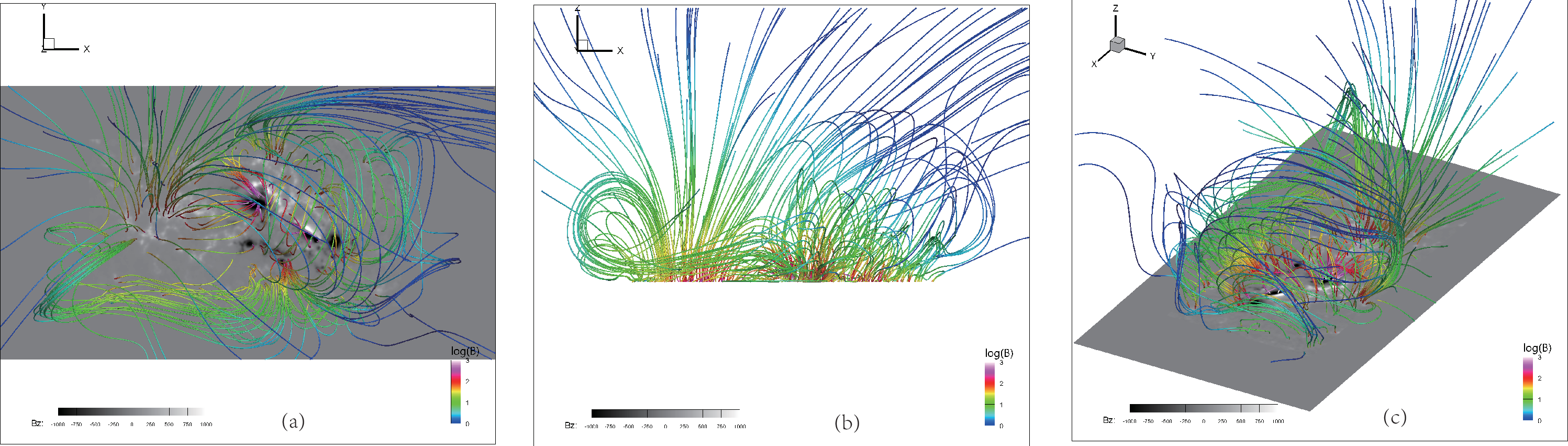}
	\caption{Non-linear force-free field extrapolation. (a)-(c) show magnetic field lines highlighting coronal funnels from different viewing angles. The right column employs height as a color bar to represent the magnetic field lines, whereas the bottom column uses magnetic field strength for color-coding. The field of view (FOV) of the magnetogram is approximately 439.5\arcsec $\times$ 235\arcsec. \label{fig:fig4}}
\end{figure}

  \begin{figure}
	\epsscale{1.2} \plotone{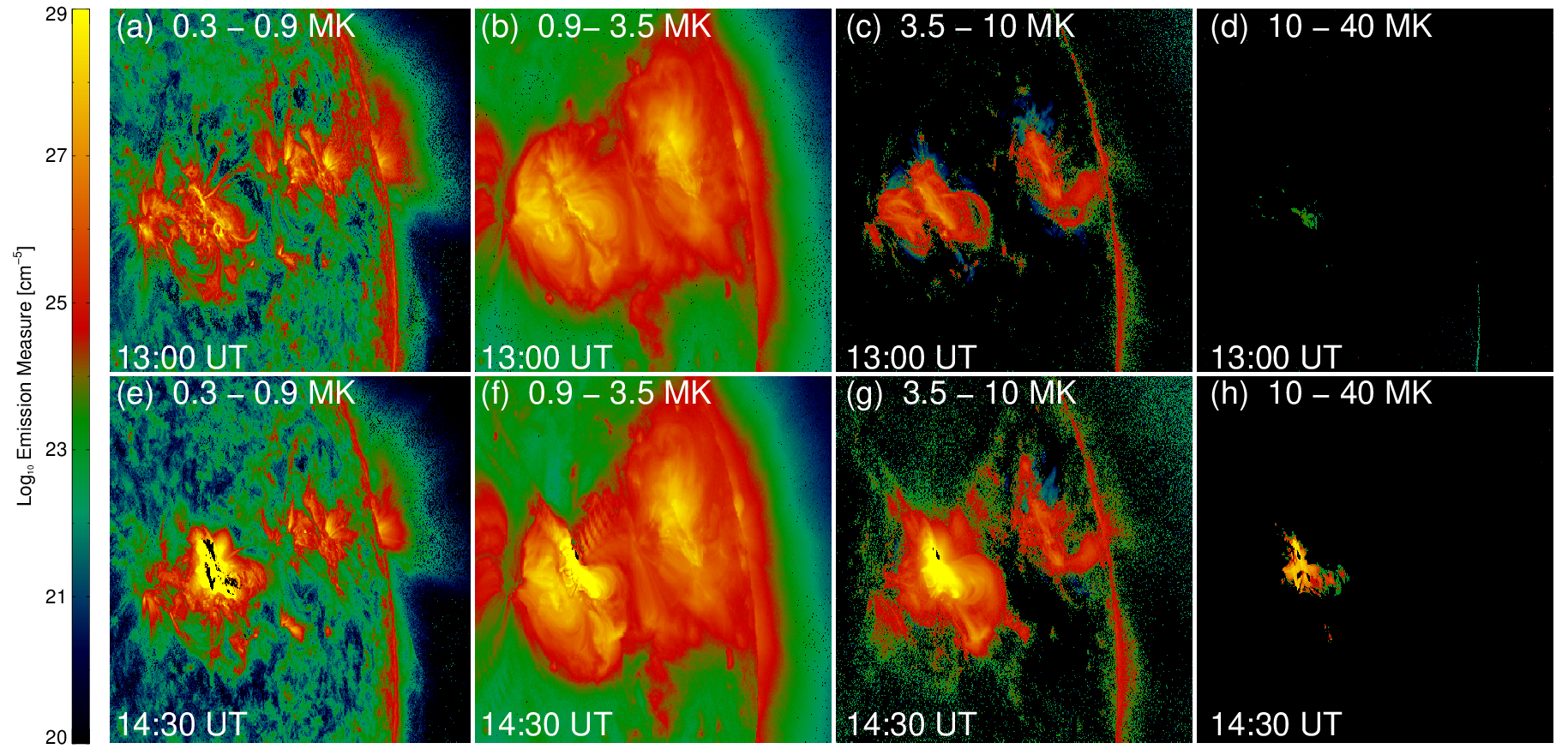}\caption{DEM maps as a function of temperatures. (a)–(d) show the DEM maps before the flare eruption at 13:00 UT. (e)–(h) present DEM maps during the QFP wave propagation at 14:30 UT.\label{fig:fig5}}
\end{figure}

In this study, an M6.0-class solar flare and the associated wave features originating from active region NOAA 11261 on August 3, 2011, are analyzed. The integrated X-ray flux in the 1\AA\ - 8\AA\ range is recorded by the Geostationary Operational Environmental Satellite (GOES). The GOES X-ray flux began to rise at 13:24 UT, peaked at 13:50 UT, and declined around 13:55 UT, exciting a pair of QFP wave trains that propagated outward along opposite directions, respectively. One of the two wave trains is a broad QFP wave, while the other is a narrow QFP wave (see Figure \ref{fig:fig1}). The flaring process and associated wave excitation are illustrated in \figref{fig:fig1} (animation.mp4)

The M6.0-class solar flare and the associated QFP wave trains are recorded by SDO/AIA. For the analysis of this event, AIA 171 \AA{} data are utilized, which are calibrated using the standard routines provided by Solar Software (SSW). Each image is normalized by its exposure time. An AIA image pixel corresponds to an angular width of $0.6\arcsec$, and the image sequences across all AIA channels have a cadence of approximately 12 seconds.

Propagation features of multi-path fast-mode QFP waves are determined through intensity distributions extracted from cuts centered at the flare site. Spatial averaging over nearby pixels is adopted to enhance observational signals. The width of the averaging area grows with distance from the flare. This technique is illustrated in \figref{fig:fig1} by a set of two sectors labeled ``A1-A2'', each with an angular extent of 10$\degree$. It should be noted that the ``A1'' slice was used to measure the kinematic characteristics of the broad QFP wave train, while the ``A2'' slice measured those of the narrow QFP wave train. These sectors were selected to sample QFP waves with significant amplitude. The spatially averaged intensities were then stacked in temporal order to create time-distance plots, as shown in \figref{fig:fig2}. To better highlight the wave propagation, running-difference images were used, where each difference image is calculated by subtracting the image taken 24 seconds earlier. Interestingly, the narrow QFP wave split into two even narrower wave trains at about 13:45 UT. Kinematic characteristics of these two wave trains are measured using the ``A3-A4'' slices, each with an angular extent of 5$\degree$, as shown in \figref{fig:fig1}.

Wavelet transforms are applied to capture periodic components from observed intensity changes at positions L1–L4 within sectors ``A1–A4'' (see \figref{fig:fig2}).
An analysis of GOES flux time derivative data was attempted in order to extract possible flare periodicities. However, time differentiation amplifies instrumental noise substantially, generating periodic peaks without adequate statistical confidence for physical interpretation. Only wavelet results derived from the L1 SDO/AIA light curve are illustrated in \figref{fig:fig3}(c), yielding a dominant periodicity of $68\pm10$ s. The narrow QFP wave propagates along the open funnel-like loops, while the broad QFP wave appears to travel through the low corona. Therefore, a nonlinear force-free field (NLFFF) extrapolation is applied to reconstruct the three dimensional magnetic structure of NOAA 11261 \citep{jiangchaowei2018,zoupeng2020}. The boundary magnetogram was provided by SDO/HMI. Magnetic field lines from strong polarities at the active region NOAA 11261 were visualized to highlight the funnel structures as displayed in \figref{fig:fig4}.

Differential emission measure inversions are obtained at 13:00 UT and 13:45 UT. The first epoch corresponds to a pre-event state, and the second corresponds to the wave propagation phase. The DEM maps at different temperature ranges are plotted in \figref{fig:fig5}. These maps reveal the plasma emissions before and during the QFP wave propagation. The thickness of the coronal funnel is estimated using the magnetic field extrapolation model for NOAA 11261. This thickness corresponds to approximately three times the typical width of a coronal loop observed in the same active region. This thickness is adopted as an estimate of the column depth $d$ of the coronal plasma. Electron density is computed based on emission measure and column depth with the relation $n_e=\sqrt{\mathrm{EM}/{d}}$.

\section{Results}
\label{sec:obs}
\subsection{Kinematics of the Bidirectional QFP Waves}

\figref{fig:fig2} presents the time-distance diagrams that show the propagation and periodic properties of the bidirectional QFP wave trains. Sectors ``A1'', ``A3'' and ``A4'' cover the multi narrow coronal waves (QFP1) and record the propagation characteristics of the narrow QFP wave trains. The propagation speeds of QFP1 are corrected for projection effects using the inclination angle of 60\degree\ relative to the plane of sky, and the deprojected speeds range from 1002 to 5032 \speed{}. The velocity uncertainty consists of two parts: a  \speed{20} fitting error from the time--distance diagram, and a projection uncertainty from the 5\degree\ inclination angle error, leading to a total relative uncertainty of $\sim$15\% for QFP1. The periods measured at locations L1, L3, and L4 are 68$\pm$10 s, 75$\pm$20 s, and 66$\pm$10 s. These periods are derived from wavelet analysis as demonstrated in \figref{fig:fig3}. Sector ``A2'' covers the broad QFP wave (QFP2). The propagation speeds of QFP2 are corrected for projection effects using the inclination angle of 10\degree, and the deprojected speeds range from 189 to 561 \speed{}. The velocity uncertainty is $\sim$10\% for QFP2, also combining the fitting error and projection error. The corresponding period is 75$\pm$20 s. All measured propagation speeds and periods are listed in Table\ref{tab:QFP}.

\subsection{Periodicities of flare emission and QFP waves}

Only statistically significant periodic signatures obtained from SDO/AIA QFP wave trains are discussed. The bidirectional QFP waves propagate along two distinctly different paths in opposite directions. Narrow and broad fast-mode wave trains show matching periodic characteristics. This similarity indicates a shared excitation source at the flare center. By contrast, periodic signatures measured from independent AIA QFP wave trains pass the 95\% significance threshold and provide robust constraints on the flare core’s energy release cycle. The consistent $\sim$66–75 s periods detected in bidirectional QFP waves indicate a unified periodic driver originating in the flaring core.

\subsection{Magnetic structure of the coronal funnels}
\label{sec:mag}

\figref{fig:fig4} displays the magnetic field lines originating from the strong magnetic polarities in NOAA 11261. The active region NOAA 11261 is located at N15$\degree$W35$\degree$ on the solar disk. The heliocentric angle from the solar disk center is calculated to be approximately 38$\degree$ according to the solar latitude and longitude. Multiple narrow QFP wave trains are guided by the well-defined funnel-like magnetic structure. The broad QFP wave propagates in the opposite direction through the lower corona. The inclination angles between the wave propagation direction and the plane of sky are estimated from the three-dimensional magnetic field geometry. The angle for the narrow QFP wave path is determined to be about 60$\degree$. The angle for the broad QFP wave path is determined to be about 10$\degree$.
The magnetic field strength is derived from the nonlinear force-free field extrapolation. For the funnel-like structure that guides the narrow QFP waves, the field strength ranges from 5 to 35 \unit{G}, which is written as $B_1^{\mathrm{extr}}=5–35\unit{G}$. For the region of the broad QFP wave, the field strength ranges from 3 to 25 \unit{G}, expressed as $B_2^{\mathrm{extr}}=3–25\unit{G}$. These values serve as the reference for verifying the seismological results.

Nonlinear force-free magnetic field extrapolation is used to estimate the transverse width of the coronal waveguides. The measured width of the magnetic funnel is about 5 Mm, which is approximately two to three times the typical width of a normal coronal loop in the active region. This value is adopted as the initial estimate of the column depth for the coronal plasma. Considering the geometric projection effect along the line of sight, the effective column depths are derived as $d_1=6.5\unit{Mm}$ for the narrow QFP wave funnel and $d_2=12.5\unit{Mm}$ for the broad QFP wave region, respectively.

\subsection{Plasma temperature and density}

\figref{fig:fig5} displays the differential emission measure (DEM) distributions covering a temperature interval from 0.3 MK to 40 MK. The flaring core is filled with hot plasma distributed over a wide range of 1–10 MK. As the QFP waves propagate outward, the flaring core is heated globally to high temperatures (see \figref{fig:fig5}(g)), which is distinctly different from the pre-flare condition (see \figref{fig:fig5}(c)). Before the flare onset, the hot plasma in the flaring core was concentrated only inside filamentary structures, as shown in \figref{fig:fig5}(d). The coronal funnel that guides the narrow QFP wave trains mainly confines plasma with temperatures in the range of 0.9 MK to 3.5 MK.

It is noted that the DEM inversion yields numerical values up to 10–40 MK, but such extremely high temperatures cannot be reliably constrained by the SDO/AIA EUV passbands. The unphysical high-temperature component is therefore discarded, with only the well-constrained DEM results below 10 MK used for quantitative analysis.

The characteristic plasma temperature is determined from the DEM peak value. For the narrow QFP wave region, the average temperature is measured as $T_1=2.0\unit{MK}$. For the broad QFP wave region, the average temperature is derived as $T_2=1.6\unit{MK}$.

The DEM is further integrated over temperature to derive the emission measure (EM). The DEM shows two well-separated peaks in the reliable range of 0.3–10 MK, corresponding to low-temperature and high-temperature plasma contributions. The high-temperature part is likely associated with low-lying coronal structures. Following the method adopted in \citet{lidong2020a}, the high-temperature component is excluded from EM calculations. The electron density is computed as $n_e=\sqrt{\mathrm{EM}/d}$, where $d$ is the line-of-sight column depth constrained by the magnetic field extrapolation. The average electron number density in the narrow QFP wave region is found to be about $5.2\times10^8\unit{cm^{-3}}$ before the flare and decreases to $4.5\times10^8\unit{cm^{-3}}$ during wave propagation. Applying the same procedure, the average electron number density in the broad QFP wave region is about $2.1\times10^8\unit{cm^{-3}}$ before the flare and reduces to $1.8\times10^8\unit{cm^{-3}}$ during the event. 

Through the DEM analysis presented in \figref{fig:fig5}(c) and (g), an obvious temperature increase in the active region is detected during the flare, indicating strong plasma heating associated with the flare energy release. The DEM results clearly show a significant enhancement of the high-temperature component, which confirms efficient plasma heating and rapid energy release during the flare eruption. This thermal variation is consistent with the enhanced multi-thermal plasma radiation revealed by the DEM, further verifying that the active region undergoes a global heating process during the flare.

\subsection{Seismological application}
\label{sec:alfen}

The observed propagation speeds span \speed{501}--\speed{2516} for the narrow QFP waves and \speed{186}--\speed{552} for the broad QFP waves, with a typical measurement uncertainty of $\pm20$\speed{} from time--distance diagram fitting. All apparent velocities are first corrected for projection effects using the inclination angles derived from the three-dimensional magnetic structure. The sound speed $C_s$ is calculated from the observational temperature, and the true Alfv\'en speed $V_A$ is obtained by $V_A=\sqrt{V_{\rm fast}^2-C_s^2}$. MHD seismology is implemented in two approaches: deriving a magnetic field interval from the full deprojected speed range, and calculating a representative magnetic field strength from the mean deprojected speed. Both sets of results are compared with nonlinear force-free field extrapolation values.

\subsubsection{Results from the full speed range}

For the narrow QFP wave with $\phi_1=60^\circ$, $T_1$=2.0 MK, the sound speed is $C_{s1} $ $\approx$ \speed{166}. With $n_1=4.5\times10^8$ cm$^{-3}$, after projection correction and subtracting the sound speed contribution, the seismological magnetic field range is $B_1$$^\mathrm{seism}$ $\approx$ 14.2$\pm2.1$--$73\pm11$ G (with $\sim15\%$ uncertainty). For the broad QFP wave with $\phi_2=10^\circ$, $T_2=1.6$ MK, the sound speed is $C_{s2} $ $\approx$ \speed{148}. With $n_2=1.8\times10^8$ cm$^{-3}$, after projection correction and subtracting the sound speed contribution, the seismological magnetic field range is $B_2$$^\mathrm{seism}$ $\approx$ 1.5$\pm0.1$--$5.2\pm0.6$ G (with $\sim6\%$--$11\%$ uncertainty).

\subsubsection{Results from the mean speed}

For the narrow QFP wave, the sound speed is $C_{s1}$$\approx$ \speed{166}. A deprojected mean speed of $v_1$=\speed{3017$\pm457$} is adopted; after subtracting the sound speed contribution, the inferred representative magnetic field strength is $B_1^\mathrm{seism}$ $\approx$ 54$\pm8$ G (with $\sim15\%$ uncertainty). For the broad QFP wave, the sound speed is $C_{s2}$ $\approx$ \speed{148}. A deprojected mean speed of $v_2$=\speed{374$\pm21$} is adopted; after subtracting the sound speed contribution, the inferred representative magnetic field strength is $B_2^\mathrm{seism}$ $\approx$ 3.4$\pm0.3$ G (with $\sim10\%$ uncertainty).

\subsubsection{Comparison and consistency}

Magnetic field strengths derived from seismological calculations after projection correction and sound speed subtraction show good consistency with extrapolation outcomes: narrow QFP features $B^\mathrm{extr}=5$--$35$ G, $B^\mathrm{seism}=14.2\pm2.1$--$73\pm11$ G (with $\sim15\%$ uncertainty), and $\langle B^\mathrm{seism}\rangle=54\pm8$ G (with $\sim15\%$ uncertainty); broad QFP features $B^\mathrm{extr}=3$--$25$ G, $B^\mathrm{seism}=1.5\pm0.1$--$5.2\pm0.6$ G (with $\sim6\%$--$11\%$ uncertainty), and $\langle B^\mathrm{seism}\rangle=3.4\pm0.3$ G (with $\sim10\%$ uncertainty). All uncertainties are propagated from the $\pm20$\speed{} fitting uncertainty in time--distance diagrams and projection geometry. All key plasma and magnetic parameters used in this study are listed in Table\ref{tab:table2}. The observed consistency confirms that multi-path fast-mode QFP events serve as effective probes for coronal magnetic field measurements.

\begin{table*}[ht]
	\centering
	\caption{Parameters of the QFP waves and flare periodicity}
	\label{tab:QFP}
\setlength{\tabcolsep}{3pt}
    \begin{tabular}{llccccccc}
\hline
		\hline

	Region &	Wave & Channel & Slit  &  Phase speed(\kms) &  Phase speed(\kms,deprojected) &  Position &  Period (s) \\
		\hline
		\multirow{3}{*}{Narrow QFP Wave Train}
	 &	QFP1 & 171 \AA & A1 & (501-1883)$\pm20$ & $1002\pm153$ -- $3766\pm570$& L1 & 68 $\pm$ 10 \\
	 &	QFP1 & 171 \AA & A3 & (2380-2516)$\pm20$   & $4760\pm720$ -- $5032\pm761$ & L3 & 75 $\pm$ 20 \\
     &	QFP1 & 171 \AA & A4 & (1357-1726)$\pm20$  & $2714\pm411$ -- $3452\pm522$ & L4 & 66 $\pm$ 10 \\
	\hline
		\multirow{1}{*}{Broad QFP Wave Train}
	 &	QFP2 & 171 \AA & A2 & (186-552)$\pm20$   & $189\pm20$ -- $561\pm22$  & L2 & 75 $\pm$ 20 \\

	\hline
		
		\hline
	\end{tabular}\end{table*}

\begin{table}
\centering
\caption{Plasma and magnetic field parameters derived from observations and MHD seismology.}
\label{tab:table2}
\begin{tabular}{lcc}
\hline
Parameter & Narrow QFP Wave & Broad QFP Wave \\
\hline
Temperature $T$ [MK] & 2.0 & 1.6 \\
Electron density $n_e$ [cm$^{-3}$] & $4.5\times10^8$ & $1.8\times10^8$ \\
Observed speed range [\speed{}] & (501--2516)$\pm$20 & (186--552)$\pm$20 \\
Mean speed $v$ [\speed{}] & 1508.5$\pm$20 & 369$\pm$20 \\
Deprojected mean speed [\speed{}] & 3017$\pm$457 & 374$\pm$21 \\
Sound speed $C_s$ [\speed{}] & 166 & 148 \\
Inclination angle $\phi$ & $60^\circ$ & $10^\circ$ \\
Column depth $d$ [Mm] & 6.5 & 12.5 \\
Extrapolated $B$ [G] & 5--35 & 3--25 \\
Seismological $B$ range [G] & 14.2$\pm2.1$--$73\pm11$ & 1.5$\pm0.1$--$5.2\pm0.6$ \\
Seismological mean $B$ [G] & 54$\pm8$ & 3.4$\pm0.3$ \\
Uncertainty & $\sim15\%$ & $\sim6\%$--$11\%$ \\
\hline
\end{tabular}
\end{table}

\section{Discussion and Conclusions}
\label{sec:sum}

A detailed analysis is provided for multi-path quasi-periodic fast-mode QFP waves linked to a flare in NOAA 11261, focusing on the coronal magnetic structure, plasma thermodynamics, and MHD seismology diagnostics. The active region is located at N15$\degree$W35$\degree$ with a heliocentric angle of approximately 38$\degree$ relative to the solar disk center, providing a favorable geometric perspective for observing the propagation characteristics of narrow and broad QFP wave components.
The nonlinear force-free field (NLFFF) extrapolation reveals that narrow QFP wave trains are strictly guided by compact magnetic funnel structures, with an extrapolated magnetic field range of 5–35 G; in contrast, the broad QFP wave propagates in the extended low corona with a weaker magnetic field range of 3–25 G. This magnetic configuration explains the distinct propagation behaviors of the two wave components: the narrow wave is confined to a dense magnetic waveguide, while the broad wave spreads freely in the diffuse coronal medium. The inclination angles between the wave propagation paths and the plane of sky are measured as 60$\degree$ for the narrow QFP wave and 10$\degree$ for the broad QFP wave, which are critical for correcting projection effects in seismological calculations.

Differential emission measure (DEM) analysis demonstrates that the flaring core undergoes global heating during QFP wave propagation, with the hot plasma distributed across a wide temperature interval of 1–10 MK. By excluding the unphysical high-temperature DEM component above 10 MK that cannot be constrained by SDO/AIA EUV passbands, reliable plasma parameters are derived: the characteristic temperature of the narrow QFP wave region is 2.0 MK, and that of the broad wave region is 1.6 MK. The electron density in the narrow wave region (pre-flare: $5.2\times10^8\unit{cm^{-3}}$, in-event: $4.5\times10^8\unit{cm^{-3}}$) is significantly higher than that in the broad wave region (pre-flare: $2.1\times10^8\unit{cm^{-3}}$, in-event: $1.8\times10^8\unit{cm^{-3}}$), consistent with the compression effect of the magnetic funnel on coronal plasma.

MHD seismology is performed using both the full deprojected speed range and the mean deprojected speed to derive magnetic field parameters, achieving robust cross-validation with NLFFF extrapolation results. The narrow QFP wave has an observed speed range of (501$-$2516)$\pm$20\speed{} (mean speed: \speed{1508.5$\pm$20}), and the corresponding seismological magnetic field range after projection correction and sound speed subtraction is 14.2$\pm$2.1$-$73$\pm$11 G (with $\sim$15\% uncertainty), with a mean value of 54$\pm$8 G (with $\sim$15\% uncertainty). The broad QFP wave has an observed speed range of (186$-$552)$\pm$20\speed{} (mean speed: \speed{369$\pm$20}), and the corresponding seismological magnetic field range after projection correction and sound speed subtraction is 1.5$\pm$0.1$-$5.2$\pm$0.6 G (with $\sim$ 6\% $-$ 11\% uncertainty), with a mean value of 3.4$\pm$0.3 G (with $\sim$10\% uncertainty).

Results highlight the importance of multi-component QFP wave analysis in constraining the spatial variation of coronal plasma and magnetic fields. The slight discrepancy between the upper limit of seismological magnetic field and extrapolation results for the narrow QFP wave can be attributed to the inherent uncertainty of NLFFF extrapolation and the rapid dynamic evolution of the coronal medium during flare eruptions. Future work will incorporate multi-wavelength observations and more sophisticated magnetic field models to reduce diagnostic uncertainties and further explore the generation mechanism of multi-path QFP waves.

This result demonstrates that strong local plasma heating occurs during the flare eruption, consistent with the overall heating process revealed by the multi-thermal structure. Consistent periodic signals extracted from dual-path QFP waves provide robust observational constraints for a unified flaring-core driver. 

Main conclusions are summarized as follows:

(1) Two distinct QFP wave components are detected in the target flare: narrow wave trains guided by magnetic funnels and a broad wave propagating in the low corona, with observed speed ranges of \speed{501–2516}and \speed{186–552}, respectively. Dynamical analysis further reveals that both fast and slow wave components appear within each of \figref{fig:fig2}(a) and (b), and the speed of the fast component is nearly three times that of the slow component, which is highly consistent with the numerical simulations of coronal MHD wave dynamics by \citet{chen2002}.

(2) Plasma diagnostics reveal higher temperature and electron density in the narrow QFP wave region compared to the broad wave region, reflecting the inhomogeneous thermodynamical properties of the flaring corona.

(3) The seismologically derived magnetic field parameters are highly consistent with NLFFF extrapolation results, demonstrating the feasibility of QFP-wave-based coronal magnetometry.

(4) Multi-path fast-mode QFP events offer strong constraints on spatial distributions of coronal magnetic fields and plasma properties.
Periodic signals derived from noisy GOES flux time derivatives are excluded from all analysis due to insufficient statistical reliability.
These events open a new way to examine dynamic coronal environments linked to solar flares.

(5) This work differs from the study by \citet{pascoe2013,pascoe2014b}, which proposed that broad QFP waves originate from the leakage of narrow QFP waves propagating along funnel-like loops. The present observations are not consistent with the leaky-wave scenario, indicating that the formation mechanisms of narrow and broad QFP waves still need to be further investigated. The narrow guided wave and broad diffuse wave components appear simultaneously and propagate along distinctly separate paths, and the broad wave is not a product of lateral leakage from the narrow guided beam. This simultaneous occurrence and independent propagation geometry cannot be explained by the leaky-wave model.

The present study finds that magnetic field results from MHD seismology match extrapolation values. This confirms that fast-mode QFP waves can be used for stable coronal magnetic field diagnosis. 

\acknowledgments

Comments from the referee are appreciated. These inputs help improve the quality of the manuscript. The SDO and STEREO teams are acknowledged for the provision of excellent observational data. Y.H.M. is supported by the National Natural Science Foundation of China (NSFC 12103016), the Fund of Shenzhen Institute of Information Technology (Nos. SZIIT2025KJ003 and HX-0951), and the High-Talent Research Funding under Grant (RC2022-001, RC2024-003). I thank Chaowei Jiang for valuable assistance with the nonlinear force-free field (NLFFF) magnetic field extrapolation. Yang Su is thanked for providing the DEM code, and C. Torrence \& G. P. Compo for making their wavelet analysis software available at \url{http://atoc.colorado.edu/research/wavelets}.

\bibliography{new.ms.bib}
\bibliographystyle{aasjournal}

\end{document}